# Deep Speckle Holography Redefines Label-free Nanoparticle Phenotyping

Yanmin Zhu[1,2,3*], Yuxing Li[1], Jingyan Chen[1], Derek Yuen-Wa Ho[4], Chutian Wang[1], Yuzhe Zhang[1], Xue-Qi Wang[5], Bo Lu[1], James Kar-Hei Fang[4,6,7], Francis Chi Chung Ling[5], Loza F. Tadesse[2,8,9*], Edmund Y. Lam[1*]

[1]Department of Electrical and Computer Engineering, The University of Hong Kong, Pokfulam, Hong Kong, 999077, China.
[2]Department of Mechanical Engineering, Massachusetts Institute of Technology, Cambridge, MA, 02139, United States.
[3]School of Biomedical Engineering, The University of Hong Kong, Pokfulam, Hong Kong, 999077, China
[4]Department of Food Science and Nutrition, The Hong Kong Polytechnic University, Hung Hom, Hong Kong, 999077, China.
[5]Department of Physics, Faculty of Science, The University of Hong Kong, Pokfulam, Hong Kong, 999077, China.
[6]State Key Laboratory of Marine Environmental Health, City University of Hong Kong, Kowloon Tong, Hong Kong, 999077, China.
[7]PolyU-BGI Joint Research Center for Genomics and Synthetic Biology in Global Ocean Resources, The Hong Kong Polytechnic University, Hung Hom, Hong Kong, 999077, China.
[8]Ragon Institute of MGH, Massachusetts Institute of Technology and Harvard University, Cambridge, MA, 02139, United States.
[9]Jameel Clinic for AI & Healthcare, Massachusetts Institute of Technology, Cambridge, MA, 02139, United States.

*Corresponding author(s). E-mail(s): ymzhu@hku.hk; elam@eee.hku.hk; lozat@mit.edu

**Nanoparticle metrology has long been constrained by the assumption that, in mixed and unprocessed fluids, particle size, morphology, composition, and species-specific abundance cannot be resolved simultaneously from a single label-free measurement. Here, we revisit this long-standing limitation by showing that complex forward speckle-holographic fields define an information-rich optical space for multidimensional particle signatures. We report deep speckle holography, a physics-informed generative framework that profiles particle identity, size, morphology, and species-resolved abundance from a single non-contact optical measurement. Across purified suspensions, mixed particle populations, environmental waters, human urine, and other unprocessed native fluids, the method enables direct nanoparticle inference without purification, labeling, or destructive preprocessing, delivering concurrent multidimensional readouts in 0.9 s over a dynamic range spanning 10 orders of magnitude. Deep speckle holography establishes a route toward direct label-free nanoparticle phenotyping in real-world fluids, moving nanoscale measurement beyond isolated-particle characterization toward multidimensional inference in complex mixtures, and expanding the scope of questions nanoscale measurement can address, from real-time tracking of nanoparticle transformations in living and environmental systems to non-invasive quality control of nanomedicine formulations, and beyond.**

For half a century, the characterization of nanoparticles (NPs) in solution has been governed by two foundational assumptions. The first is the purity prerequisite, in that accurate physicochemical measurement demands rigorous sample purification, such as, dilution, filtration, labeling, or physical isolation, to suppress the confounding influence of heterogeneous backgrounds [1-5]. The second is the one-dimension-at-a-time principle, which assumes that distinct NP attributes (size, morphology, composition, surface chemistry, and abundance) must be interrogated by disparate, specialized instruments [6-9]. Together, these assumptions have shaped the prevailing metrological framework, in which NP phenotyping depends on prior isolation and stepwise interrogation of individual properties.

A profound and rarely acknowledged consequence is that, because existing techniques rely on either physical separation or ensemble averaging, knowledge of NPs in their native environments is derived almost exclusively from highly processed, idealized aliquots. The act of measurement itself perturbs the measurand. Sample preparation alters aggregation states [10], strips protein coronas [11,12], disrupts concentration gradients [13,14], and eliminates the very inter-particle interactions that govern NP behavior *in situ* [15,16]. In effect, the field has been characterizing artifacts of its own methodology rather than the NPs themselves.

The fragility of this orthodox framework becomes starkly apparent in complex, multi-component matrices. Conventional optical methods operate under the ensemble-averaging paradigm, which implicitly treats the superposition of scattering signals from heterogeneous NP populations as an irreversible loss of single-particle identity [17,18]. Multiple scattering, in particular, has been regarded as intractable noise that must be suppressed through extreme dilution. Single-particle techniques such as nanoparticle tracking analysis circumvent ensemble averaging but impose their own purity prerequisite [19]. They require optically transparent, low-concentration suspensions and cannot efficiently resolve species-specific abundance in unknown mixtures. Single-particle mass spectrometry offers elemental sensitivity but is inherently destructive, limited to inorganic compositions, and dependent on meticulous calibration [9, 20]. The field has thus resigned itself to a fundamental trade-off between non-destructiveness and single-particle specificity. No existing approach simultaneously provides label-free, multiparametric, single-particle-resolved physicochemical phenotyping directly in unprocessed complex fluids.

Here, we present evidence that these foundational assumptions are neither physically inevitable nor experimentally necessary. We demonstrate that multiple scattering in a dense, heterogeneous NP suspension does not result in irreversible information loss. Instead, it generates a highly deterministic, spatially coherent interference field. This holographic speckle fingerprint encodes the multiparametric physicochemical identities of every constituent species. What the field has long dismissed as noise is, in fact, an extraordinarily rich data stream awaiting the correct profiler.

To retrieve these latent physicochemical signatures, we introduce deep speckle holography (DSH), a generative artificial intelligence-assisted speckle holographic platform. Unlike conventional approaches that operate within the constraints of purity preprocessing and single-dimension interrogation, DSH simultaneously abandons both limitations by exploiting the full spatially coherent scattering field as a complete physicochemical signature source. On the hardware side, a compact, lensless dynamic forward speckle holographic configuration captures the full spatially coherent speckle-holographic composite field in a single sub-second frame. This process encodes particle size, morphology, complex refractive index, and inter-particle scattering correlations directly into the light field. Algorithmically, DSH embeds a dynamic forward speckle holographic model as a physics prior into a multimodal generative network. The system jointly analyzes local interference textures and global optical features to invert the superposed scattering field into species-resolved, multiparametric physicochemical phenotypes. This physics-informed framework accelerates convergence and ensures fine-grained sensitivity to complex scattering signals while preserving interpretability across an extensive range of NP sizes and abundance levels.

To validate this paradigm reversal, we systematically dismantle each assumption of the orthodox framework. We demonstrate that DSH discriminates isodiametric but compositionally distinct particles, refuting the claim that ensemble optical methods cannot resolve material identity. The system deconvolves bimodal and multimodal particle-size mixtures with exceptional fidelity, overturning the irreversible information loss assumption. It achieves species-resolved abundance quantification in stress-test systems of 70 nm diamond and 100 nm gold, extending single-particle phenotyping to the regime where conventional holography fails. Finally, DSH operates directly in unfiltered seawater, wastewater effluent, and human urine, abolishing the purity prerequisite entirely. To our knowledge, this is the first demonstration of simultaneous, label-free phenotyping of size, morphology, composition, and species-resolved abundance from mixed NP populations directly in unprocessed biological and environmental fluids. This capability shifts nanometrology from purified samples toward direct, multidimensional analysis in native complex media.

# Results

## Recasting coherent scattering from nuisance to nanoparticle information field

The orthodox framework dictates that as nanoparticle abundance increases, the transition from single scattering to multiple scattering destroys individual particle signatures. Consequently, multiple scattering is universally treated as intractable noise that must be suppressed through physical dilution. DSH fundamentally inverts this assumption by treating the superposed coherent scattering field not as degraded information but as a deterministic optical manifestation of the complete physicochemical state of the suspension.

When a coherent monochromatic beam passes through an unpurified NP suspension, each particle functions as a complex optical mask. The incident wavefront undergoes particle-specific phase shifts and amplitude modulations. At low abundance, this produces isolated in-line holographic fringes. In dense systems, these single scattering events cascade into complex multiple scattering and interference trajectories. Rather than yielding stochastic noise, this coherent superposition generates a highly structured speckle holographic field. This field continuously imprints the refractive index, morphology, size, and spatial distribution of every constituent species directly onto the propagating light.

This regime is commonly interpreted as a breakdown of conventional single-particle interpretability. However, our measurements show that the composite field remains stable under fixed conditions and varies systematically with particle properties and suspension composition. These observations suggest that the transition to optical complexity does not remove nanoparticle information, but redistributes it into a higher-order interference pattern that is not captured by ensemble-averaged readouts.

To analyze this regime, DSH couples lensless dynamic forward speckle holography with a physics-informed multimodal generative framework (Fig. 1A, B). The imaging configuration records the full spatially coherent composite field in a single sub-second exposure without purification, labeling, or optical isolation. A forward speckle–holographic model is incorporated as a physics prior, allowing local interference textures and global optical features to be analyzed jointly. This framework enables the species-specific, multiparametric physicochemical information to be resolved from the superposed scattering field.

We assessed this approach across purified suspensions containing polymeric, inorganic, metallic, and carbon-based particles, as well as environmental waters and human urine. Across diverse monodisperse and mixed populations including polystyrene, poly(methyl methacrylate), poly(lactic-co-glycolic) acid, titanium dioxide. gold, and diamond NPs (Fig. 1C), the framework achieves > 90% analytical fidelity, root mean square error remained below 0.28, $R^2$ exceeded 0.95, and robust coefficient of variation remained below 5.58%. Under varying laser intensities and extreme background noise levels, the system simultaneously resolves particle identity, size, morphology, and species-resolved abundance directly from a single sub-second optical frame. Together, these results indicate that increasing optical complexity in heterogeneous NP suspensions does not necessarily erase NP information, but can preserve it in a form that remains amenable to computational inference.

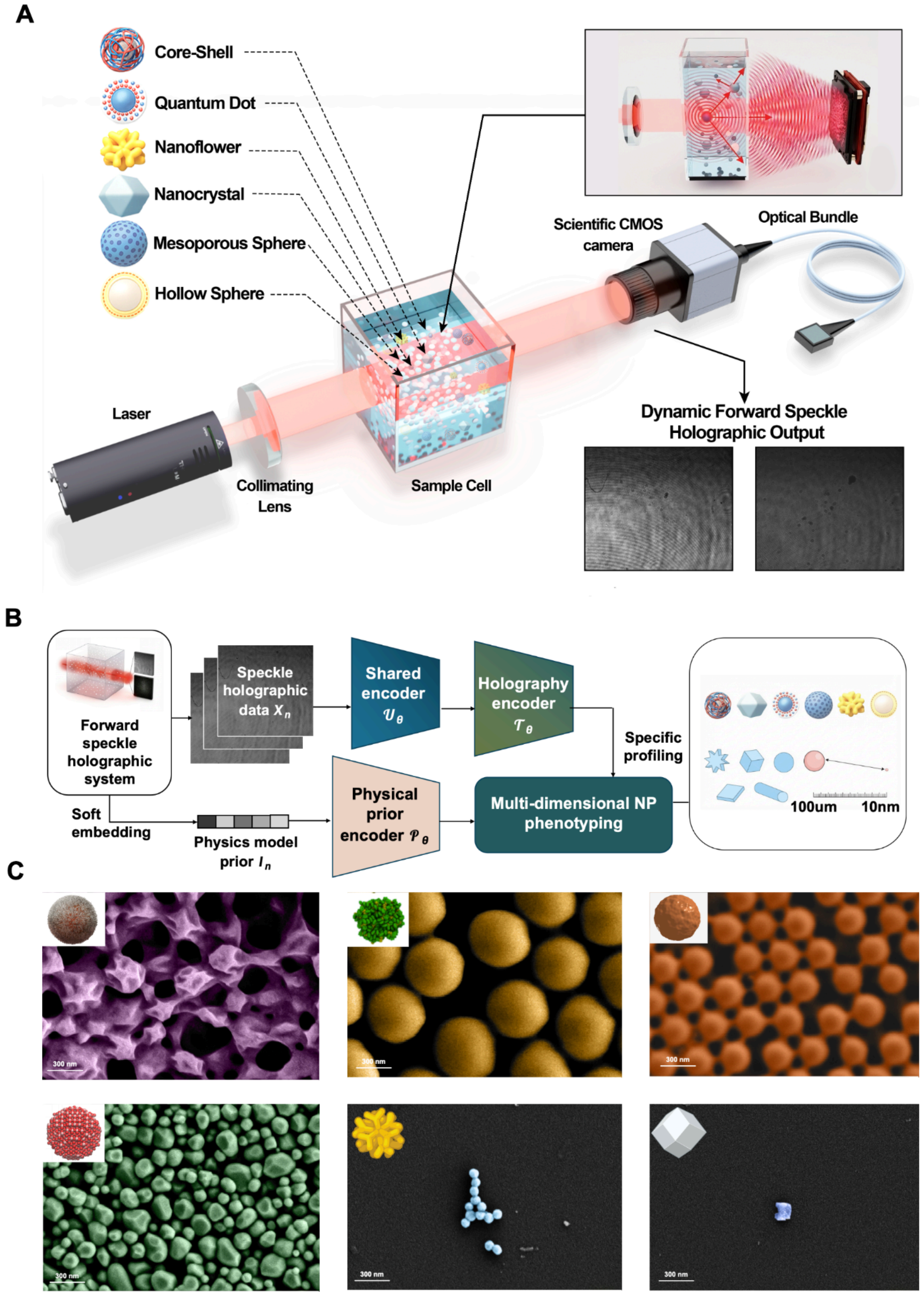

**Fig. 1 DSH treats superposed coherent scattering as an analyzable information-bearing field and leverages an optical-computational framework to resolve NP-specific information across diverse systems.**

**(A)** Optical configuration for capturing the dynamic forward speckle holographic field. This superposed coherent field $\langle I \star I \rangle = c_m (S \star S) * (h_z \star h_z)$ continuously imprints the material dependent scattering structure $(S \star S)$, complex refractive index, and the propagation-dependent blur kernel $(h_z \star h_z)$, and encodes the NP signatures. Subplot at the top right: optical principle description of the dynamic forward speckle holographic imaging field. During transmission, the plane wave deforms into an "object wave" encoding complex refractive index $(n_m = n_r + jn_i)$ and topographical features, interfering with transmitted and scattered waves, generate holographic fringes. As NP abundance increases, single scattering events transition into dense multiple scattering cascades. Instead of destroying information, this transition generates a deterministic speckle dominated field that inscribes the multidimensional physicochemical signatures of the entire interacting population. **(B)** Architecture of the physics-informed multimodal generative network. By embedding the forward holographic formation model as a soft physical prior, the network establishes a shared representation space that fuses local interference textures with global scattering anchor points. This framework resolves the superposed coherent field to achieve simultaneous multiparametric phenotyping without requiring physical sample purification. A two-stage training strategy with physics-informed constraints prevents overfitting and ensures robust resolving. **(C)** Scanning electron microscopy characterizations of representative NP systems alongside their structural models. The evaluated systems include 500 nm poly(lactic-co-glycolic acid), 500 nm polystyrene, 300 nm polymethyl methacrylate, 300 nm titanium dioxide, 70 nm gold, and 70 nm diamond NPs. Furthermore, the experimental validation encompasses isodiametric but compositionally distinct systems, heterogeneous particle size mixtures, extreme stress test groups, and unprocessed biological and environmental matrices. This comprehensive evaluation confirms the universality of the DSH across diverse physical regimes.

## Preservation of particle-specific signatures within complex scattering fields

We first establish a performance baseline by phenotyping single-component NP solutions, a regime where conventional ensemble methods are expected to perform adequately. Even in this idealized scenario, DSH achieves quantification fidelity that matches or exceeds established techniques across nine NP preparations spanning 70 nm to 5 μm and abundance levels from 0.005 to 10 mg/mL (Fig. 2).

Specifically, Fig. 2B displays dynamic forward speckle holograms of 500 nm PMMA at abundance levels from 0.01 mg/mL to 25 mg/mL. At lower particle abundance levels (e.g., 0.01 mg/mL), individual holographic interference patterns in the solution can be observed. As the particle abundance levels increases, the frequency of multiple scattering and interference rises. At 25 mg/mL, multiple-scattering speckle patterns dominate the forward dynamic speckle holograms. Fig. 2C presents the abundance profiling results for 500 nm PS, 500 nm $TiO_2$, 300 nm PMMA, 500 nm PLGA, 100 nm gold, and 70 nm diamond, respectively. From the scatter plots, all measured abundancedistribute around the red dotted line representing $y = x$. Among these, 500 nm PS (yellow), 500 nm $TiO_2$ (green), and 500 nm PLGA (pink) exhibit the closest alignment with the true abundance level line, with their error distribution histograms staying very close to zero. The estimation for 300 nm PMMA are slightly more dispersed but remain tightly concentrated around the true abundance levels. While its error histogram shows a slightly larger spread, most errors still fluctuate narrowly around zero. The measurement distributions of 100 nm gold and 70 nm diamond are broader than that of 300 nm PMMA, with 70 nm diamond showing higher dispersion. Smaller particle sizes can increase measurement error because their reduced scattering cross-section leads to weaker and less distinct holographic interference patterns, making their individual optical signatures more challenging to resolve against background noise. Moreover, 100 nm gold and 70 nm diamond particles each have approximately a 10% diameter variation in manufacturing, which affects the abundance estimates. By contrast, the manufacturing diameter variation for 500 nm PS, $TiO_2$, PMMA, and PLGA is below 5%.

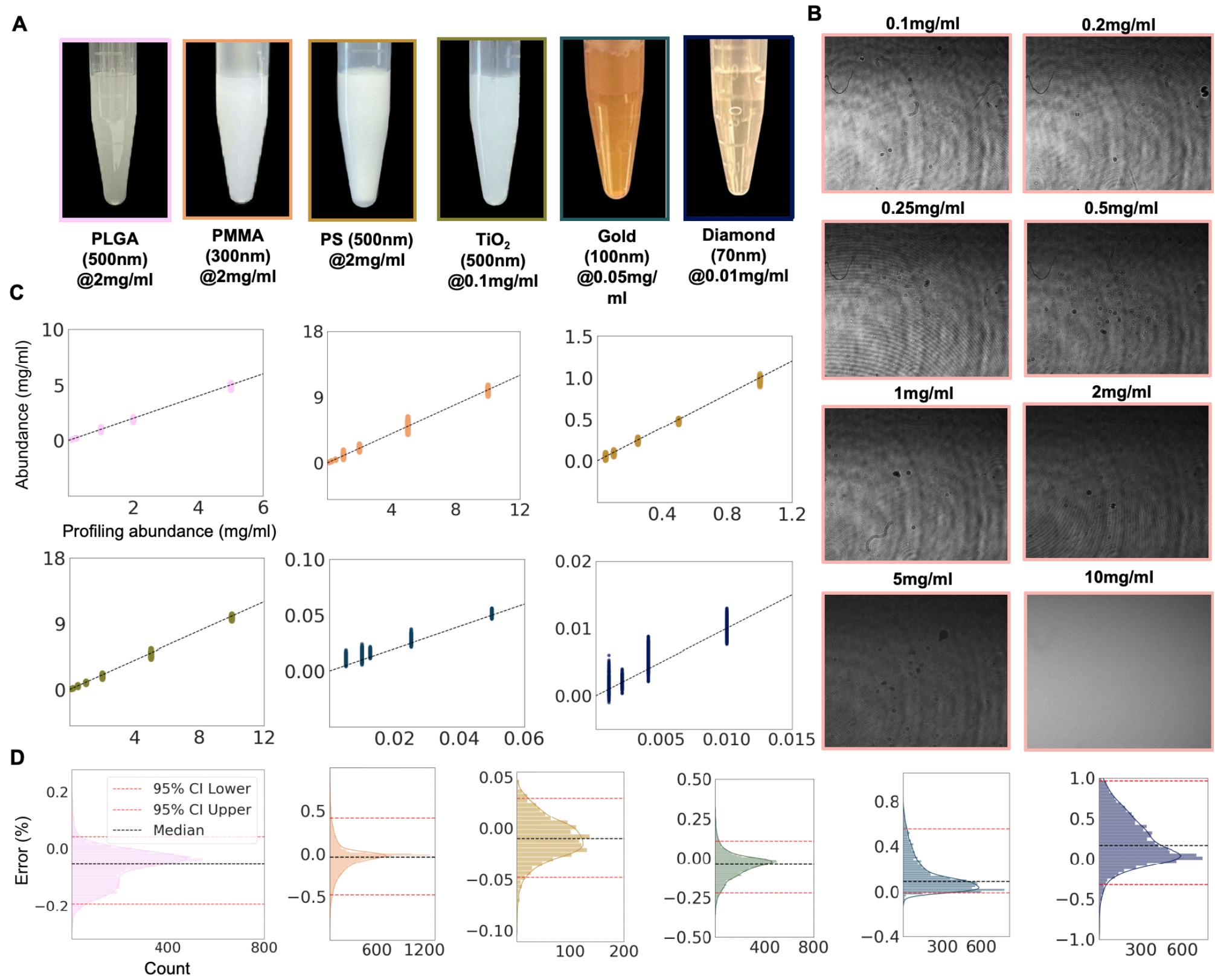

**Fig. 2 Quantitative resolving of particle-specific signatures from coherent scattering fields in monodisperse systems.**
**(A)** Representative images of the evaluated single particle suspensions. **(B)** Dynamic forward speckle holographic data for polymethyl methacrylate NPs across abundance levels from 0.01 to 25 mg/mL. As particle abundance increases, individual holographic interference patterns transition into a multiple scattering dominated speckle field. **(C)** Species-resolved abundance quantification fidelity across six distinct NP populations. The scatter plots demonstrate the alignment between the computationally retrieved abundance and the ground truth values where the dashed line represents perfect agreement. (**D**) Error distribution histograms with 95% confidence interval (CI) for the abundance retrievals. The tight clustering around zero confirms the high analytical fidelity of the DSH across diverse material compositions and particle sizes.

From the mean error ($\mu$) and standard deviation (σ) shown in the supplemental methods S3.4 and Table S3.4, we observe the following: for 500 nm PS in the 0.1-1 mg/mL range, $\mu$ is 0.0092 mg/mL, and σ is 0.0197 mg/mL. Within 0.1–10 mg/mL for 500 nm $TiO_2$ and 300 nm PMMA, the mean errors μ are 0.0963 mg/mL and 0.0061 mg/mL, and σ of 0.150 mg/mL and 0.252 mg/mL respectively. For 500 nm PLGA, the $\mu$ is 0.0826 mg/mL and the σ is 0.0952 mg/mL. At lower abundance levels (0.001 – 0.050 mg/mL), 100 nm gold has a $\mu$ of 0.0033 mg/mL and a σ of 0.0024 mg/mL. In the very low abundance range (0.001 – 0.01 mg/mL), 70 nm diamond exhibits a $\mu$ of 0.0008 mg/mL and a σ of 0.001 mg/mL. For all the experimental groups, when we actively introduce the noise during the experiments, e.g., 500 nm $TiO_2$, 300 nm PMMA, and 500 nm PS, a slightly high σ are shown, which demonstratesthe high sensitivity of the DSH. Overall, these results demonstrate our method's capability to achieve accurate NP signature unveiling over a broad range of particle diameters (tens of nanometers to hundreds of nanometers) and abundance levels (0.001–10 mg/mL). In Fig. 3A, we list the test results for nine types of NPs at 16 different abundance levels. We calculated the ratio of each measured result to the true NP abundance level to evaluate the phenotyping fidelity. The experimental results demonstrate consistently high phenotyping fidelities across all tested particles. The fidelity remains over 93.3% for 500 nm PLGA, stands above 92.7% for 500 nm PMMA, surpasses 91.7% for 300 nm PMMA, stays above 92.7% for 500 nm PS, remains higher than 93.3% for 500 nm $TiO_2$, exceeds 91.7% for 200 nm $TiO_2$, stays beyond 89.5% for 100 nm gold, and remains above 87.6% for 70 nm diamond. Among them, 100 nm gold and 70 nm diamond serve as "stress test" groups, given their extremely small particle diameters, low abundance levels, and 10% particle size fluctuation range. Overall, the phenotyping fidelity across the system remains around or above 90%.

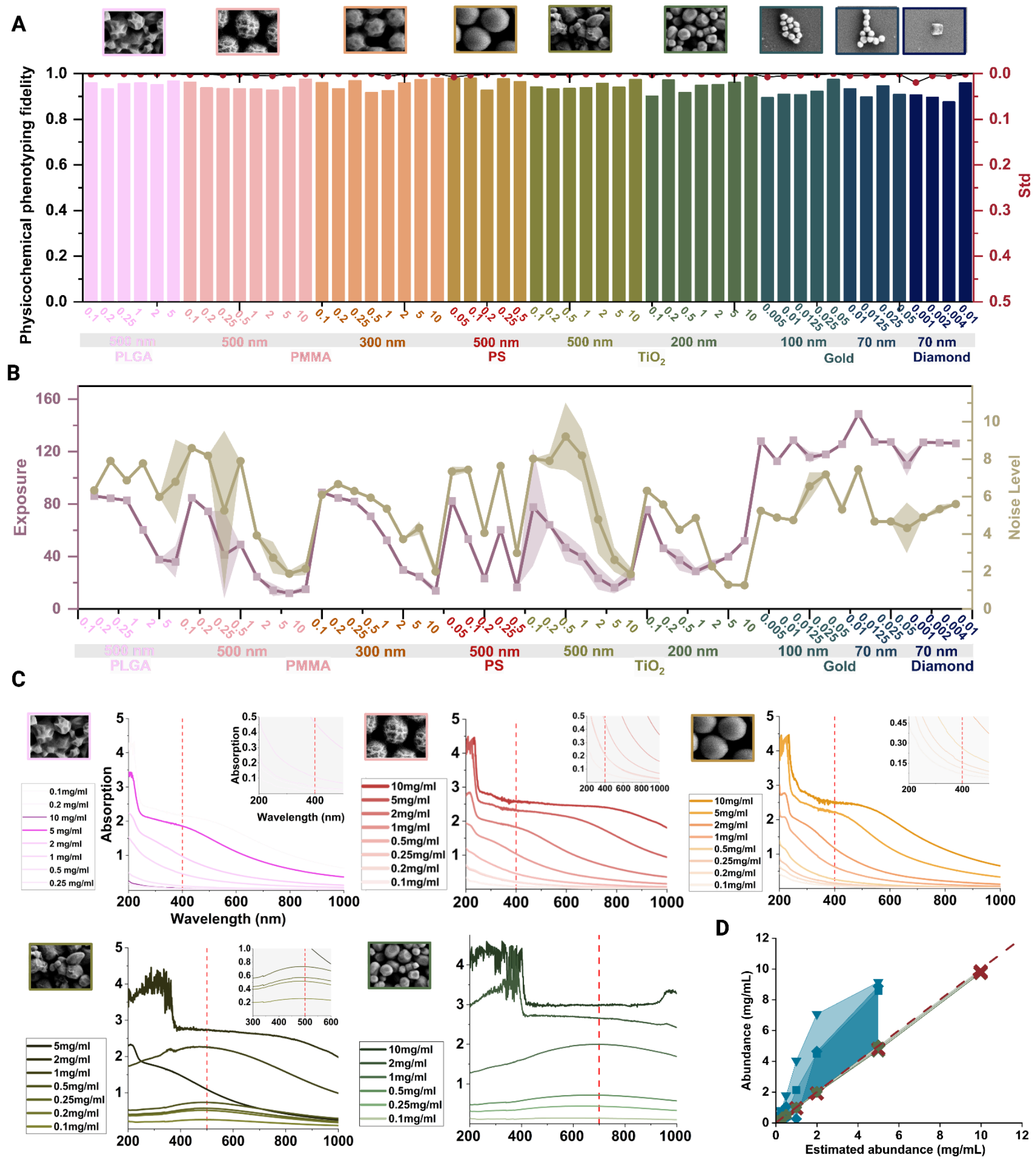

**Fig. 3 Analytical robustness of DSH across diverse physical regimes.**

(**A**) Physicochemical phenotyping accuracy (left y-axis) and standard deviation (right y-axis) across various single NP systems. The analytical fidelity consistently exceeds 94% across diverse materials including extreme stress test groups of 100 nm gold and 70 nm diamond particles. (**B**) System stability evaluation under varying optical exposure (left y-axis) and environmental noise levels (right y-axis). The framework maintains robust phenotyping performance despite significant intergroup and intragroup fluctuations in illumination and noise interference. **(C)** Conventional ensemble averaged readouts obtained via UV-Vis spectroscopy for comparison. Absorbance saturation occurs at higher abundance levels indicating the fundamental dynamic range limitations of ensemble methods. From left to right and top to down: 500 nm PLGA, 500 nm PMMA, 300 nm PMMA, 500 nm $TiO_2$, and 200 nm $TiO_2$. (**D**) Performance comparison between DSH and UV-Vis spectroscopy. DSH surpasses the ensemble averaging limit by delivering higher analytical fidelity and a significantly broader dynamic range.

We then examined the stability of the system and method under varying light intensities. As shown in Fig. 3, when testing the 500 nm PLGA, 500 nm PMMA, 300 nm PMMA, 500 nm PS, 500 nm $TiO_2$, and 200 nm $TiO_2$, the light intensities of each group differ significantly ranging from 8 to 150 (a.u.), resulting in considerable inter-group variations. We also conducted extreme-case tests, such as a 0.25 mg/mL 500 nm PMMA solution and a 0.1 mg/mL 500 nm $TiO_2$ solution, where within-group fluctuations are bigger than 70 and 90 in normalized intensity, respectively. Even with these fluctuations, the method still achieved over 90% profiling fidelity and demonstrated excellent robustness.

Additionally, we tested the noise tolerance of the system and method, measuring noise levels, defined as the standard deviation of high-frequency image components obtained after subtracting a Gaussian-blurred grayscale image from the original image, for each NP at each abundance level. This definition capitalizes on the typical manifestation of image noise as high-frequency random fluctuations, rendering the standard deviation of the high-frequency residual a robust metric for noise intensity [21, 22]. Please see Method 4.8 and Supplemental Notes S6 for details. The results are shown in Fig. 3B right y-axis and indicate that under low-noise conditions (e.g., 5 mg/mL and 10 mg/mL 200 nm $TiO_2$ solutions), high-noise conditions (e.g., 0.1 mg/mL and 0.2 mg/mL 500 nm PMMA solutions), and large noise fluctuations (e.g., 5 mg/mL 500 nm PLGA, 0.25 mg/mL 500 nm PMMA, 2 mg/mL 500 nm $TiO_2$, and 0.001 mg/mL 70 nm diamond solutions), the profiling fidelity of specific NPs remained above or approached 90%. The experimental results indicate that this method places slight demands on the noise level of the collected data, maintaining reliable measurement performance under both low- and high-noise conditions, including situations with significant noise fluctuations. We further assessed the wide dynamic range detection performance of DSH, with results presented in Fig. S3.5.1. These data demonstrate that DSH achieves a wide dynamic detection range of up to 4.4 decades, accompanied by an average profiling fidelity of 92.4%. Notably, nearly all individual profiling fidelity of specific NPs is within either the “excellent” range (> 95%) or “good” range (90%−95%), underscoring the consistency and reliability of its performance across the entire dynamic range.

We next analyzed the cumulative distribution function (CDF) of the profiling fidelity of DSH, with results shown in Fig. S3.5.2. This CDF analysis reveals that over 82.1% of samples achieve a profiling fidelity exceeding 90%, and nearly 40% of samples reach a profiling fidelity of over 95%. These statistics further present that DSH maintains high profiling fidelity across a large majority of test cases, reinforcing its robustness for practical detection scenarios. In Fig. S3.5.3, we display the abundance level-response curve of DSH, which quantifies the relationship between NP abundance level and DSH’s detection performance. It specifically illustrates average accuracies and their associated variances across different NP abundance levels. Error bars in the curve represent standard deviations of the measured accuracies, while the values of $n$ denote the number of independent testing groups for each abundance level point, ensuring transparency in experimental replication. Notably, across the entire tested abundance range, DSH maintains consistently high profiling fidelity with minimal variance, directly demonstrating its strong robustness to abundance level fluctuations in practical testing scenarios. These data show that increasing optical complexity does not monotonically destroy NP information. Instead, the representation of that information shifts from isolated fringe-like signatures toward higher-order field organization, which remains resolvable by DSH.

## Surpassing the limitation of ensemble averaging phenotyping

The ensemble averaging paradigm, exemplified by UV-Vis spectroscopy, represents the dominant logic of conventional nanometrology. Under this paradigm, the superposition of scattering and absorption signals from heterogeneous particle populations is treated as an irreversible loss of species-specific identity, and the measurement output is reduced to a single scalar metric. We selected UV-Vis spectroscopy as the closest available proxy for comparison (see Method 4.4 for method principle and operations), not as a competing technique, but as a representative of the foundational assumptions that DSH overturns. The absorption measurement results are shown in Fig. 3C.

Specifically, we used UV-Vis spectrometer to measure solutions, from left to right and top to down, of 500 nm PLGA NP solutions, 500 nm PMMA NP solutions, 500nm PS NP solutions, 500 nm $TiO_2$ NP solutions, and 200 nm $TiO_2$ NP solutions at various abundance levels. The absorbance of the 200 nm $TiO_2$ NP solution becomes saturated starting at the abundance levels of 2 mg/mL, indicating the maximum measurable range. Similarly saturation also existed in the 500 nm $TiO_2$ NP solution at 2 mg/mL and above. In Fig. 3C, we have included magnified plots of the results for the lower-abundance groups, helping to show the clear absorbance measurements for those sets. Fig. 3D summarizes the measurement data for all five NP solutions. We calculated the absorbance at 700 nm wavelength to indirectly estimateabundance (indicated by the red dashed line) using the Beer-Lambert Law [23, 24], which was relatively stable and clearly observable. The results are displayed in Fig. 3D and listed in Table S1.1. The red dashed line denotes the regression line corresponding to the true abundance. The blue symbols indicate measurements obtained using the UV-Vis spectrometry, with the blue-shaded region showing the discrepancy between the UV-Vis results and the true abundance values. The green symbols denote the measurements from our method, and the green-shaded area shows the discrepancy between our results and the true values. Red “×” markers represent data points lying outside the measurement range of UV-Vis.

From these results, it can be seen that the UV-Vis spectrometer measurements exhibit relatively large errors in the full range. In multiple samples, such as 1 mg/mL of 500 nm PLGA, 500 nm PMMA, and 2 mg/mL of 200 nm $TiO_2$, 500 nm PLGA, and 300 nm PMMA solutions, the measurement error higher than 100% of their true abundance level. For the 2 mg/mL 500 nm PMMA NP solution, the error even surpasses 200% of their true abundance value. The UV-Vis spectroscopy measurements have relatively small errors for abundance levels in the 0.1-1 mg/ml range. In comparison, our method achieves results that closely match the actual solution abundances over the entire range of 0.1-10 mg/mL for all five types of NPs tested. Our method exhibits a wide measurement range, high phenotypingg and profiling fidelity, and broad applicability for a variety of NPs. Moreover, the UV-Vis spectrometer requires a 15-minute warm-up, and each single-abundance level measurement takes approximately 5-7 minutes based on the absorbance integration time. The detailed steps for the sample preparation and UV-Vis operation are shown in Supplemental Method S2. In contrast, our method requires no warm-up period, and each single-abundance level data acquisition only takes 900 ms, demonstrating that our method significantly improves on the measurement efficiency compared to UV-Vis spectrometer.

## Simultaneous species-resolved phenotyping in heterogeneous NP mixtures

The orthodox framework asserts that when multiple NP species coexist in suspension, their overlapping scattering signals produce an irreversible loss of species-specific identity. We directly test and refute this assumption. We prepared binary mixtures of same-material but different-diameter NPs, alongside mixtures of same-diameter but different-material NPs. These systems are macroscopically indistinguishable (Fig. 4A and 4b), and no existing single optical technique can directly resolve them into species-specific abundances.

Specifically, we examined $TiO_2$ NPs at 200 nm and 500 nm in the following respective species-resolved abundance levels: 6 mg/ml and 2 mg/ml and 4 mg/ml and 4 mg/ml. We also investigated PMMA NPs at 500 nm and 300 nm in the following respective species-resolved abundance levels: 2.5 mg/ml and 7.5 mg/ml and 5 mg/ml and 5 mg/ml. DSH deconvolves the superposed holographic speckle field into species-resolved abundance estimates with fidelity up to 98.36%, with all measurements exceeding 91.94% (Fig. 4C). During the phenotyping, both low and high noise levels (Fig. 4D, Test Group 4 and Group 3), as well as small and large noise fluctuations (Fig. 4D, Test Group 1 and Test Group 4) have been included and tested. The system and method demonstrated stable phenotyping performance across all scenarios. Moreover, as shown in Fig. 4C, our system and method maintain stable performance even when ambient lighting and the light source intensity fluctuate significantly, demonstrating strong robustness against light intensity variations.

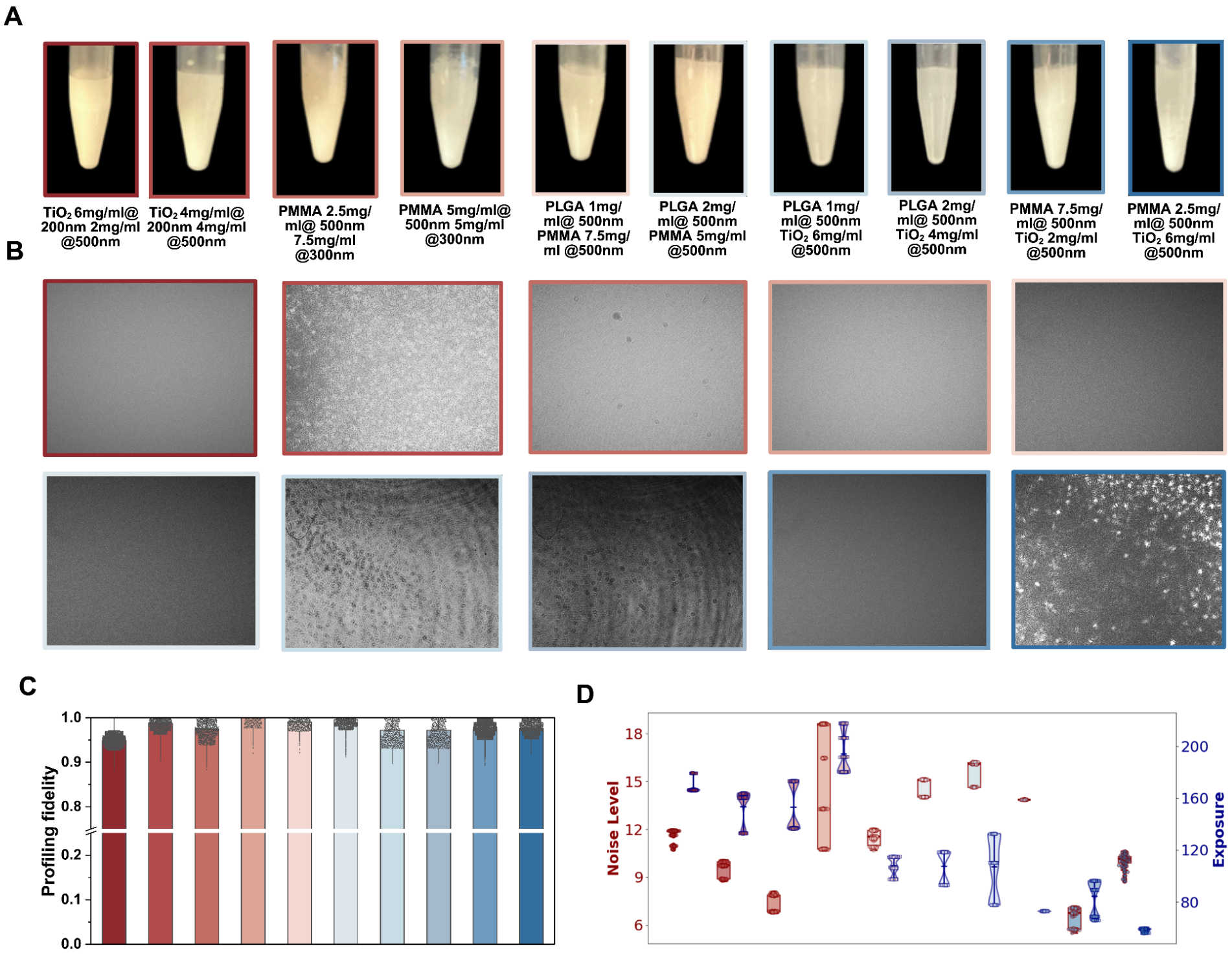

**Fig. 4 Simultaneous species-specific resolving in heterogeneous multicomponent mixtures.**

(**A**) Composition of the evaluated heterogeneous mixtures including varying diameters of $TiO_2$ and PMMA alongside cross material mixtures of diverse polymeric and inorganic particles. From left to right: 200 nm and 500 nm $TiO_2$ NPs at 6 mg/ml and 2 mg/ml and 4 mg/ml and 4 mg/ml, 500 nm and 300 nm PMMA NPs at 2.5 mg/ml and 7.5 mg/ml and 5 mg/ml and 5 mg/ml. Mixed NPs materials systems with PLGA and PMMA NPs at 1 mg/ml and 7.5 mg/ml and 2 mg/ml and 5 mg/ml. PLGA and $TiO_2$ NPs at 1 mg/ml and 6 mg/ml and 2 mg/ml and 4 mg/ml. $TiO_2$ and PMMA NPs at 2 mg/ml and 7.5 mg/ml and 6 mg/ml and 2.5 mg/ml. (**B**) Representative dynamic forward speckle holographic data for each heterogeneous mixture. Data sets are color coded to correspond with the sample categories illustrated in panel A. (**C**) Particle-specific abundance retrieval accuracies for the complex mixtures. The framework successfully resolves the superposed scattering fields to achieve species-resolved quantification with accuracies exceeding 91%. (**D**) Evaluation of analytical stability against environmental perturbations. Box plots and violin plots illustrate the diverse noise and exposure levels across experimental groups confirming robust multiparametric phenotyping under fluctuating optical conditions.

We also measured complex mixed-solution systems containing different types of NPs of the same particle diameter (500 nm). For example, we examined PLGA and PMMA NPs at 1 mg/ml and 7.5 mg/ml, and 2 mg/ml and 5 mg/ml respectively. PLGA and $TiO_2$ NPs at 1 mg/ml and 6 mg/ml, and 2 mg/ml and 4 mg/ml respectively; $TiO_2$ and PMMA NPs at 2 mg/ml and 7.5 mg/ml, and 4 mg/ml and 5 mg/ml respectively. As shown in Fig. 4C, DSH demonstrated stable high phenotyping fidelities of specific NPs, achieving up to 98.36%, with all fidelities exceeding 91.94%. Furthermore, as disclosed in Fig. 4D, the abundance profiling fidelity varied by only ~ 3% despite a 46% variance in noise interference and a 49% change in light intensity. This demonstrates that DSH maintains robust performance against light intensity fluctuations, regardless of whether ambient light and source intensity are high or low. Based on dynamic forward speckle holography and a physics-informed multimodal generative model, this approach accurately profiles species-resolved abundances in complex NP mixture systems and simultaneously resolves variations in NP diameter and material composition.

## Moving NP inference and phenotyping beyond the purified-sample paradigm

Having demonstrated that species-resolved phenotyping is achievable in controlled laboratory mixtures, we now confront the most deeply entrenched assumption in nanometrology, the purity prerequisite. We deployed DSH directly in raw, unfiltered open-water environmental samples collected from sewage treatment works and coastal sites in Hong Kong (Fig. 5A). These matrices contain unpredictable populations of environmentally weathered particles spanning diverse materials, morphologies, and size ranges (Fig. 5B). No sample purification, dilution, or labeling was performed.

Despite this extreme heterogeneity, DSH achieved abundance quantification fidelity consistently above 94% for spiked 500 nm PMMA NPs (Fig. 5E). This demonstrates that the necessity of sample purification is an instrumental artifact of previous methods, not a physical inevitability. Based on the optical photothermal infrared (O-PTIR) spectroscopy (Fig. 5D and S7–16) examination results, no 500 nm PMMA NPs were detected in any of the open-water environmental samples. We then introduced 500 nm PMMA NPs at a abundance level of 0.25 mg/mL into each environmental sample and profiled their abundances under interference from other complex environmental particles in the solution. Figs. 5C demonstrate the dynamic forward speckle holographic data of each sample and illustrate that various types, abundances, and unknown scales of particles were present in the open-water environmental samples, potentially interfering with the 500 nm PMMA NP measurement. However, the results show that the abundance level fidelity for the 500 nm PMMA NPs was consistently above 94%, with small standard deviations. As shown in Fig. 5F, distinguishable noise level fluctuations and exposure changes can be identified among each group, however the fidalities remain over 94% for all test groups.

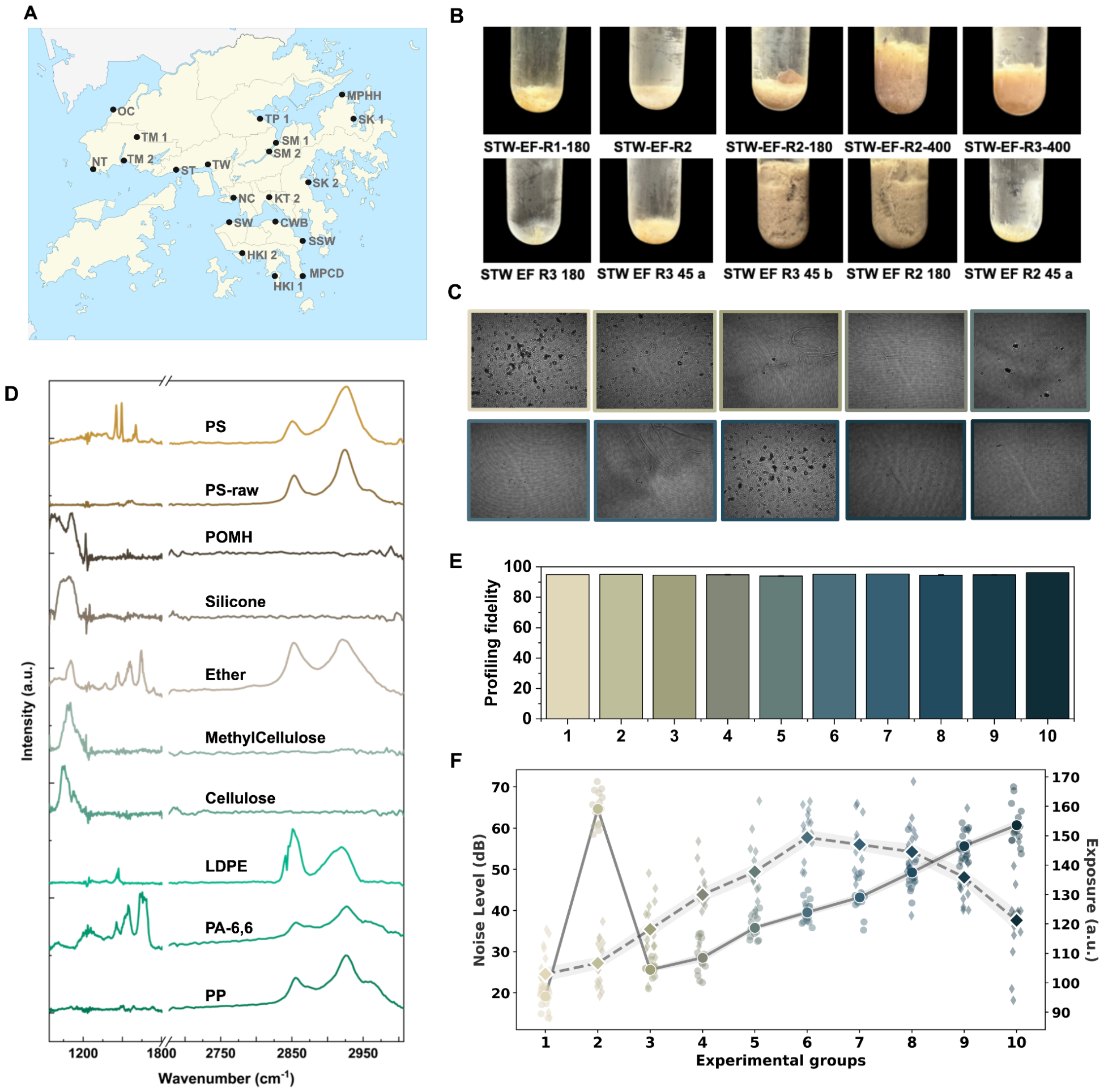

**Fig. 5 Direct label-free NP phenotyping in unprocessed environmental matrices.**

(**A**) Geographical mapping of the open water environmental sample collection sites across diverse coastal and wastewater treatment locations. (**B**) Characterization of the native environmental matrices revealing a highly heterogeneous background of unpredictable environmental particles and biological debris.. (**C**) Recorded dynamic forward speckle holographic data of the unprocessed environmental samples demonstrating extreme optical complexity. (**D**) O-PTIR spectra confirming the baseline composition of the native environmental samples. Species-specific abundance retrieval for target NPs. The framework maintains a phenotyping fidelity of 92.94% without requiring physical isolation or sample purification. (**F**) Quantification of inherent noise and intensity fluctuations across the environmental samples further validating the robustness of DSH.

## Biological fluids reveal an *in situ* regime inaccessible to conventional preprocessing-dependent metrology

To probe the limits of *in situ* phenotyping, we test DSH on human urine, which is a biological matrix of exceptional complexity containing proteins, metabolites, salts, and endogenous micro- and NPs of diverse polymeric compositions (Fig. 6A, B). O-PTIR spectroscopy confirmed the presence of cellulose, polyethylene, polystyrene, polycarbonate, polyamide, silicone, and other polymeric species across the four urine samples (Fig. 6C).

Specifically, sample 1 (UR01) contained cellulose. Sample 2 (UR02) contained cellulose, methyl cellulose, polyethylene, polystyrene, silicone, and sodium carboxymethyl cellulose. Sample 3 (UR03) contained cellulose, polyamide-6,6, polycarbonate, polyethylene, polypropylene, polystyrene, silicone, and sodium carboxymethyl cellulose. Sample 4 (UR04) contained cellulose and methyl cellulose. Without any sample preparation, DSH resolved particle physicochemical signatures and quantified particle abundance with fidelity exceeding 92% (Fig. 6D). This result abolishes the purity prerequisite for NP phenotyping in biological fluids and opens a direct path toward clinical nanoscale diagnostics.  Based on the dynamic forward speckle holographic data (as sample demonstrated in Fig. 6B), our method unveiled particle signatures and profiled particle abundances in the urine samples with accuracies exceeding 92% among various particles (Fig. 6D), with high noise level and exposure fluctuations as illustrated in Fig. 6E.

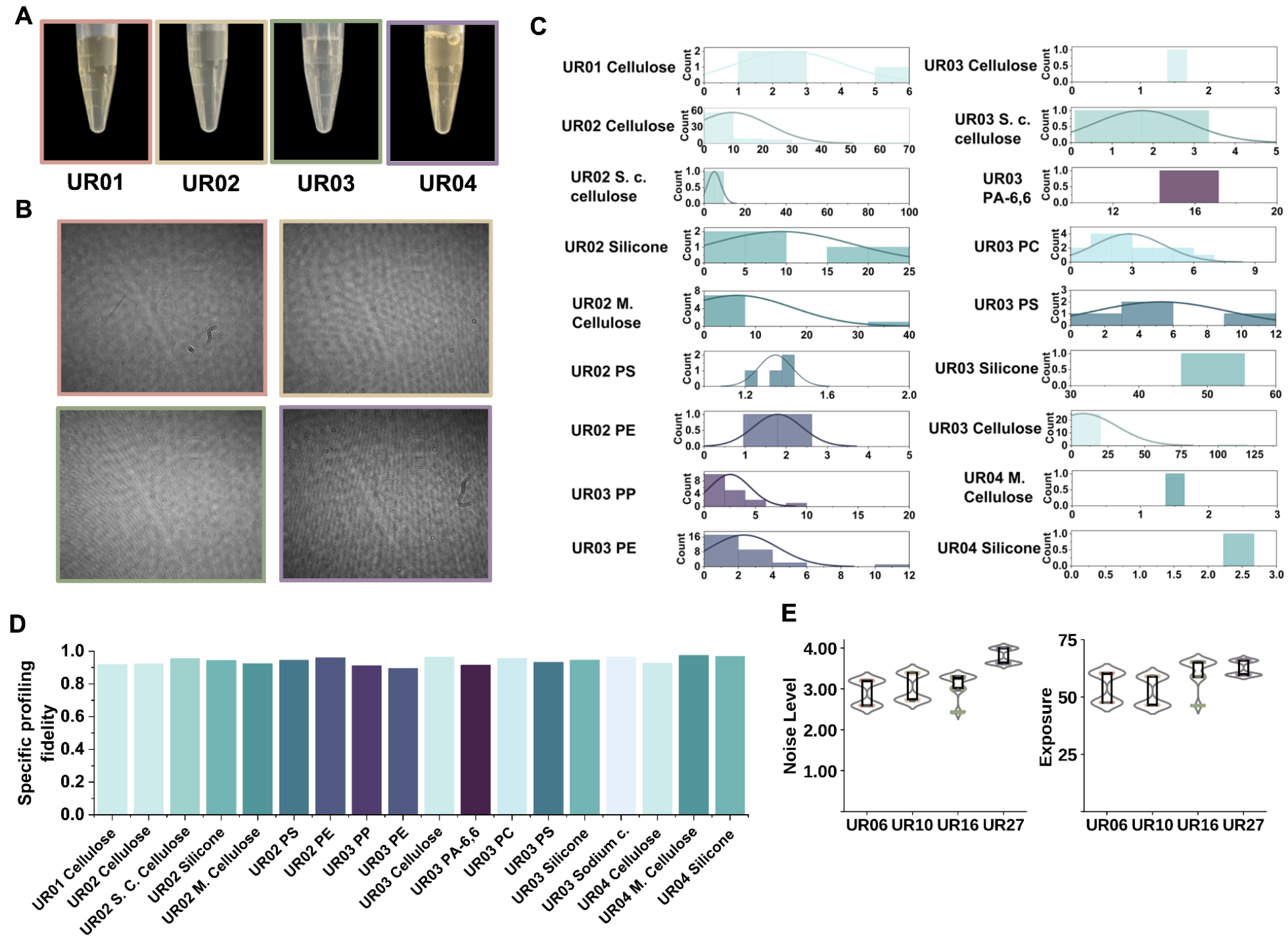

**Fig. 6** ***In situ*** **multiparametric profiling in complex biological fluids.**

**(A)** Photographic representation of the unprocessed human urine samples utilized for direct optical interrogation. (**B)** Dynamic forward speckle holographic data acquired directly from the raw biological fluids. Data are color coded to match the sample designations in panel A. (**C)** Compositional analysis of the endogenous microscale and nanoscale particles natively present within the urine samples. (**D)** Species-specific abundance quantification fidelity within the biological matrices. The framework achieves measurement accuracies exceeding 90% directly in raw urine, thereby abolishing the purity prerequisite for clinical nanoscale diagnostics. Abundance quantification fidelity for urine samples exceeds 90.91%. **(E)** Presentation of the noise and exposure variations during the unprocessed biological samples phenotyping.

## Discussion

The results presented here resolve a longstanding and seemingly fundamental paradox in nanometrology concerning the mutually exclusive relationship between *in situ* observation and multiparametric specificity. For decades, the field has accepted that non-destructive optical methods are inherently limited to ensemble-averaged, single-dimensional readouts, while multiparametric, species-resolved characterization requires destructive sample processing and sequential interrogation by multiple specialized instruments [1, 2, 25, 26]. DSH dissolves this dichotomy by demonstrating that a single coherent optical field, captured in a sub-second exposure, contains sufficient information to simultaneously resolve size, morphology, composition, and species-specific abundance, provided the field is resolved through a physics-informed computational framework rather than reduced to a scalar absorbance or intensity metric.

The key conceptual advance is the reframing of multiple scattering and interference of the light. Where the orthodox framework treats multiple scattering as an information-destroying process that must be suppressed [15, 17, 26-28], DSH treats it as an information-encoding process that must be exploited. Each scattering, reflection, refraction, and inference event imprints material-specific phase shifts and amplitude modulations onto the propagating wavefront. The resulting speckle-holographic pattern is not random noise but a deterministic, high-dimensional profiling of the suspension's complete physicochemical state. The physics-informed multimodal generative network serves as the resolver to interrogate the interpretable phenotypic parameters. This is not an incremental improvement in measurement sensitivity or range. It is a redirection of the field's foundational logic from signal suppression to signal exploitation.

This paradigm shift resolves the contradictions inherent in all existing approaches. Ensemble methods, such as UV-Vis spectroscopy and dynamic light scattering, suffer from irreversible signal superposition, collapsing the identities of heterogeneous particle populations into a single averaged metric [15, 29]. Single-particle techniques, such as NP tracking analysis and single-particle inductively coupled plasma mass spectrometry [9, 19, 30], circumvent ensemble averaging but impose their own purity prerequisite, requiring optically transparent, low-abundance suspensions and destructive sample processing. The field has thus resigned itself to a fundamental trade-off between non-destructiveness and single-particle specificity. DSH navigates this impasse by achieving single-particle-resolved phenotyping without sacrificing the non-destructive and high-throughput capability of ensemble optical methods. Our experimental results across purified suspensions, controlled mixtures, environmental waters, and human urine confirm that the one-dimension-at-a-time principle and the purity prerequisite are instrumental artifacts of previous methods, not physical inevitabilities.

By redefining the boundaries of what can be measured *in situ*, DSH does not merely answer existing questions more efficiently. It generates new classes of questions that were previously unformulable within the orthodox framework. Several testable hypotheses emerge directly from this work. First, the real-time evolution of morphological and compositional signatures during protein corona acquisition in human biofluids [31-33] might be accessible through continuous optical interrogation, enabling longitudinal monitoring of corona formation without perturbing the process itself. Second, the transient intermediate states of NP aggregation, degradation, and photoaging in environmental ecosystems, which have remained invisible to destructive sampling approaches [34-36], could, in principle, be captured through time-resolved analysis of coherent scattering fields, thereby providing direct kinetic data for environmental fate modeling.

These questions demand a fundamental recalibration of the research agenda in nanometrology and applied nanoscience. The field could pivot away from developing increasingly complex purification protocols for *ex situ* analysis and, to be suggested, invest in deploying computational optical frameworks to track NP life cycles directly within their native habitats. The modular architecture of DSH, in which the physics-informed profiler is fixed while the task-specific estimator is interchangeable, provides a scalable foundation for this transition. As laser sources [37], imaging sensors [38-41], and generative network architectures [42] continue to advance, this platform is positioned to enable rapid, on-site, and potentially wearable NP phenotyping while maintaining high fidelity and sensitivity. By providing a universal and label-free platform for multidimensional phenotyping, DSH empowers researchers to interrogate the fundamental behaviors of nanomaterials in precision medicine, environmental toxicology, and advanced manufacturing under truly representative conditions.

# Method

## Dynamic forward speckle holographic imaging field

When a coherent laser beam is incident on an NP solution, the complex electric field arriving at the camera (e.g., CMOS sensor) is the coherent superposition of a large number of propagation paths $E_k$ [43],

$$E_{tot}(r) = E_0(r) + \sum_i E_i^{(1)}(r) + \sum_{i,j} E_{ij}^{(2)}(r) + \sum_{i,j,k} E_{ijk}^{(3)}(r) + \ldots, \quad (1)$$

where $E_0$ represents the ballistic (direct) light, $E_i^{(1)}$ is light that has been scattered once by the i-th particle, and $E_{ij}^{(2)}$, $E_{ijk}^{(3)}$, . . . , represent all higher-order paths that involve multiple times scattering.

Specifically, during the ballistic light $E_0$, the photon traverses the sample without being scattered. This field retains a plane wave and acts as a reference lightwave for interference with all scattered components. In the single-scattered field, $E_i^{(1)}$, the photon is deflected only once by particle $i$ before reaching the detector. Interference between $E_i^{(1)}$ and the ballistic wave produces in-line holograms. In multiple-scattered fields, $E_{ij}^{(2)}$, $E_{ijk}^{(3)}$, . . . , the photon undergoes two or more scatterings before exiting the cuvette. Multiple scattering continuously alters the optical path intensity and phase, causing the patterns to appear as random speckle or a more uniform diffuse background on the imaging plane.

The intensity that the camera records is

$$I(r) = |E_{tot}(r)|^2, \quad (2)$$

where all scattering orders and the interferences between them are encoded into the pattern. The recorded pattern is jointly determined by all the above paths.

To quantitatively model the optical signature of each material, we assume that scattering only occurs at the particle sites, which locally modulate the amplitude and phase of the incident wave according to their refractive index and thickness [44]. Each NP is characterized by a complex refractive index $n_m = n_r + jn_i$, with $n_r$ accounting for refractivity and $n_i$ for absorptivity relative to the surrounding medium. Let $c_m$ denote the areal abundance (particles per unit area) of this material in the suspension. In the thin-sample regime, where light primarily undergoes single scattering, the transmitted complex field can be modeled as a spatially modulated wavefront:

$$S(x) = a(x)exp(j\Phi(x)), \text{ with } \Phi(x) = \frac{2\pi}{\lambda} n_m \omega(x), \quad (3)$$

where $\omega(x)$ is the local thickness profile of the suspended particles, and $a(x)$ accounts for their absorption. This field serves as the input to the coherent forward model that governs the final speckle intensity distribution.

Using Fourier optics [28], the total order of scattering process of a single particle can be approximated as a free-space propagation process as that in inline holography, which is determined by

$$E_{tot}(x) \approx S(x) * h_z(x), \quad (4)$$

where $h_z(x) = h(x; z)$ is the impulse response for the propagation distance $z$.

Applying the convolution–correlation identity for coherent systems yields

$$I \star I = (S * h_z) \star (S * h_z) = (S \star S) * (h_z \star h_z). \quad (5)$$

Although the scattered field from a single nanoparticle is fully deterministic, the total scattered speckle field is stochastic. For a fixed material class with complex refractive index $n_m$ and areal abundance $c_m := N/A$, we model the total transmitted field as

$$S_{tot}(x) = \frac{c_m}{N} \sum_{i=1}^{N} S(x - x_i), \quad (6)$$

where $S(x)$ is the complex field from a single nanoparticle and $x_i$ denotes the location of the $i$-th particle. Since the particle positions are random, $S_{tot}(x)$ tends toward a complex circular Gaussian distribution.

Using statistical optics in [45], we derived the intensity autocorrelation under deterministic forward propagation. Taking the ensemble average gives:

$$< I \star I >= c_m (S \star S) * (h_z \star h_z), \quad (7)$$

and we provide an explicit derivation in the Supplemental Method S3.1.

Our system is designed to profile the multidimensional physicochemical signatures of particles across a wide range of radii and materials, such as particles with a diameter of 70 nm - 100 μm. With a 532 nm illumination wavelength, both Rayleigh scattering and Mie scattering may occur during imaging. The single scattering model is not suitable for the description of the scattered light field. In this system, each particle contributes to the self-referenced interference between the scattered spherical wave and the ballistic light beam to form in-line holographic fringes. As the particle abundance level increases, the single-particle holographic fringes overlap each other, and the fringing visibility gradually decreases (Fig. 2B).

In the experiments, a FU650AD100-PXG2698-20mm laser (Fuzhe Laser Technology)

with a wavelength of 532 nm and a chip power of 100 mW was used. The overall dimensions of the laser module are 19×26×98 mm. The beam spot produces a 20 mm diameter circular beam. A beam collimation lens (diameter 22.0 mm, center thickness 7.0 mm, focal length 60 mm) is placed in front of the light source. The imaging sensor is a Crevis MG-A500P-22 (monochrome) with a resolution of 2464 × 2056 and a pixel size of 3.45 µm × 3.45 µm.

## Physics-informed multimodal speckle holographic generative network

This model adopts a multi-task learning paradigm that softly integrates prior knowledge from dynamic forward speckle holographic imaging into the deep learning training process through a triple mechanism of explicit input-level incorporation, task-sequential design, and physics-based learning objectives. This ensures that the model not only achieves data-level fitting but also maintains reasonable consistency between the physics model and the training results.

In stage-I, for the nth raw speckle-holographic image $X_n$ and the physics prior $I_n$, an encoder $f_\theta$ is trained to learn multi-dimensional signatures of the NPs:

$$h_n = f_\theta(X_n, I_n). \tag{8}$$

Subsequently, we train a multimodal physical mapper $g_\phi$ that learns to compress complex scattering observation data into a set of standardized physical descriptors:

$$\hat{s}_n = g_\phi(h_n), \qquad \sum_{i=1} \hat{s}_{n,i} = 1. \tag{9}$$

Then, we optimize the representation loss to guide the network in learning feature representations of different physical models in the speckle holographic system, enabling the network to extract key information related to physical quantities from the complex speckle holographic textures

$$L_{rep} = -\frac{1}{N}\sum_{n-1}^{N}\sum_{i} t_{n,i} ln\hat{s}_{n,i}, \tag{10}$$

where $t_{n,i}$ is the target feature distribution.

In stage-II, we freeze the parameters of the trained encoder $f_\theta$ and train an estimator $h_\psi$ to predict the particle-sensitive quantification values

$$\hat{c}_n = h_\psi(f_\theta(X_n, I_n)) \tag{11}$$

Then, we optimize the quantification loss between the prediction values and the ground-truth particle-sensitive quantification values

$$L_{qt} = \frac{1}{2N}\sum_{n=1}^{N} ||\hat{c}_n - c_n^{*}||, \tag{12}$$

where $N$ is the batch size.

In total, the network in DSH aims to optimize

$$\mathcal{L}_{\text{total}} = \begin{cases} \mathcal{L}_{\text{rep}}(\mathbf{X}, \mathbf{I}; \theta, \phi), & \text{stage} - \text{I} \\ \mathcal{L}_{\text{qt}}(\mathbf{X}, \mathbf{I}; \theta^{*}, \psi), & \text{stage} - \text{II} \end{cases} \tag{13}$$

where $\theta^{*}$ represents the fixed encoder weights in stage-I.

This physics soft-embedding approach helps the deep learning network to compensate for the inherent limitations of traditional physical models. Since NP solution systems are in a real-time dynamic state, their forward dynamic speckle holographic fields are changing, involving multi-scale random scattering, interference, absorption, and diffraction propagation. However, theoretical physical models cannot comprehensively consider and fully simulate the uncertainty factors in image acquisition and imaging systems, such as camera noise, particle oxidation [46], morphological and characteristic changes, imaging distortion, temperature fluctuations, Brownian motion randomness [47], and other complex environmental factors. Consequently, hard-coding an explicit model (i.e., "hard embedding") has limitations [48-50] and may not adequately support networks in accurately performing precise measurements based on multi-dimensional signatures of NPs (See Supplemental Method S4.2 and Supplemental Figs. S6-S8). The soft-embedding approach can compensate for the unmodelable deviations and experimental condition variations through data-driven learning, enhancing the model's adaptability and robustness to real complex environments. Furthermore, soft-embedding alleviates the need for precise calibration of system parameters under hard model constraints, which is often resource-intensive and error-prone. By reducing such dependencies, it facilitates real-time, high-throughput experimental measurements and implementations, thereby enhancing the method's practical applicability in diverse settings and *in situ* measurements.

The two-stage network structure and training strategy further support network migration, allowing reuse across various NP signature-sensitive tasks (e.g., particle size distribution estimation). When switching tasks, only the stage-II network structure needs adjustment, providing flexibility and convenience for quick repurposing without overhauling the entire framework. This modular design minimizes retraining overhead and supports scalable deployment in multi-task scenarios.

In this method, the multimodal network input takes the physics prior as a physical anchor, and image features capture the spatial distribution information that theory struggles to fully model. The combination preserves interpretability while enhancing profiling fidelity.

The network training and testing are computed by one NVIDIA T4 GPU and one 2.3 GHz Quad-Core Intel Core i5 CPU. For details on network parameters and structure, please refer to the Supplemental Method S4.2.

## Materials and Reagents

PMMA NPs (300 nm and 500 nm) were purchased from Tomicro Tech. at an initial abundance of 25 mg/mL. PS (500 nm) and PLGA (500 nm) NPs were purchased from Zhichuan Tech. at an initial abundance level of 10 mg/mL. $TiO_2$ (200 nm and 500 nm) were purchased from Shuangying Materials Tech. Co.. Gold NPs (70 nm and 100 nm) were purchased from Langfei Biotechnology Co. with an initial abundance of 0.05 mg/mL. Diamond NPs (70 nm) were obtained from Adamas Nanotechnologies, Inc. at an initial abundance of 0.01 mg/mL.

Nine monodisperse NP preparations constituted the panel used for network training experiments: 500 nm PLGA, 300 nm and 500 nm PMMA, 500 nm PS, 200 nm and 500 nm $TiO_2$, 70 nm and 100 nm gold (Au) and 70 nm diamond. Reported diameters correspond to the manufacturers' intensity-weighted hydrodynamic sizes. All procedures were carried out in a class-II biosafety cabinet with sterile, endotoxin-free consumables. Liquids were transferred with calibrated single-channel air-displacement micropipettes (Eppendorf, Hamburg, Germany), and all dilutions were brought to a final volume of 1.00 mL with ultrapure water (18.2 MΩ cm, Milli-Q). Immediately after preparation every suspension was vortex-mixed for 30 s and bath-sonicated (40 kHz, 5 min; Branson 2510) and used for imaging within 1 h to minimize agglomeration.

PLGA was supplied as a 10 mg/mL aqueous stock. Aliquots of 10, 20, 25, 100, 200 and 500 µL were dispensed into 1.5 mL low-binding polypropylene tubes and adjusted to 1.00 mL, yielding working abundance levels of 0.10, 0.20, 0.25, 1.0, 2.0 and 5.0 mg/mL. PMMA (25 mg/mL) was available in two diameters (300 nm and 500 nm). For each diameter, 4, 8, 10, 20, 40, 80, 200 and 400 µL of stock were diluted to 1.00 mL, producing 0.10, 0.20, 0.25, 0.50, 1.0, 2.0, 5.0 and 10 mg/mL. The PS stock (500 nm, 25 mg/mL) was portioned at 2, 4, 8, 10 and 20 µL and made up to 1.00 mL to give 0.05, 0.10, 0.20, 0.25 and 0.50 mg/mL. $TiO_2$ dispersions (20 mg/mL, 200 nm and 500 nm) were diluted by transferring 5, 10, 25, 50, 100, 250 and 500 µL into separate tubes and bringing each to 1.00 mL, affording 0.10, 0.20, 0.50, 1.0, 2.0, 5.0 and 10 mg/mL. Lotspecific gold NP stocks (70 nm and 100 nm) were diluted to obtain 0.005, 0.010, 0.0125, 0.025 and 0.050 mg/mL; the required volumes were calculated from the certificateof-analysis abundance levels. The ND dispersion (70 nm, 0.020 mg/mL) was diluted by transferring 50, 100, 200 and 500 µL and adjusting to 1.00 mL, resulting in 0.001, 0.002, 0.004 and 0.010 mg/mL. To improve sample diversity and evaluate model robustness, binary mixtures were prepared at three volumetric ratios—3 : 1, 1 : 1 and 1 :3—while maintaining a total volume of 1.00 mL. Five diameter- or material-matched combinations were generated: (i) 500 nm PLGA and 500 nm PMMA, (ii) 500 nm $TiO_2$ and 500 nm PLGA, (iii) 500 nm $TiO_2$ and 500 nm PMMA, (iv) 200 nm $TiO_2$ and 500 nm $TiO_2$ and (v) 300 nm PMMA and 500 nm PMMA. As an illustration, for the mixed $TiO_2$ solution, 100 µL of the 200 nm stock plus 300 µL of the 500 nm stock (3 : 1), 200 µL with 200 µL (1 : 1) and 300 µL with 100 µL (1 : 3) were each diluted to 1.00 mL. All mixed suspensions were vortexed and sonicated as described above before downstream processing.

Open-water environmental samples were collected from sewage treatment works and effluent at the sites disclosed in the Fig. 5A Samples STW-EF-R1-180, STW-EF-R1-C, STW-EF-R2-180, and STW-EF-R3-180 are filtered by filters with a 180 µm pore size. Samples STW-EF-R2-45a, STW-EF-R2-45b, STWEF-R3-45a, and STW-EF-R3-45b are filtered by filters with a 45 µm pore size. Samples STW-EF-R2-400 and STW-EF-R3-400 are filtered by filters with a 400 µm pore size. Samples are retained on the mesh that was rinsed and preserved in ethanol and stored at 4◦C. See Supplemental Method S8 for detailed sample collection protocol.

Urine samples were collected from four adult volunteers (18–65 years of age) who were recruited via advertisements at the University of Hong Kong and surrounding communities. All volunteers provided written informed consent prior to their participation in the study.

## UV-Vis spectrometry

NPs fabricated from certain materials can interact strongly with light at specific wavelengths. Their characteristic extinction peaks in the UV–Vis region can be used to quantify NP population abundance level in solution using the Beer–Lambert law [23, 24,

51]:

$$A = \epsilon D c \quad (14)$$

where $A$ is the absorbance, $\epsilon$ is the molar extinction coefficient (in $M^{-1} \cdot cm^{-1}$), $D$ is the path length (in cm), and $c$ is the NP population abundance level in the solution. For the same materials of NPs, the $\epsilon$ is identical. Therefore, the NP population abundance value in a solution can be estimated from repeated measurements of $A$, following the equation:

$$c = \frac{A}{\epsilon D} \quad (15)$$

For the same experimental setup, $D$ remains constant.

Aqueous solutions of $TiO_2$ (200, 500 nm), PLGA, and PMMA (300, 500 nm) NPs were prepared in deionized water at various abundance levels (from 0.1 to 10 mg/mL), homogenized via vortex mixing (30 s), and analyzed using a UV-Vis spectrophotometer (MAPADA, UV-3200). Measurements were conducted in a 1cm × 1cm quartz cuvette, with parameters set to a wavelength range of 200−1000 nm (high-to-low scan), 1 nm data interval, medium scan speed (~ 240 nm/min), and 2 nm slit width. Prior to sample analysis, a de-ionized (DI) water blank was measured for baseline correction. The cuvette was rinsed with DI water and sample solution step by step, then dried with lint-free tissue, finally filled to ~ 80% capacity for each measurement, ensuring optical alignment and prevention of cross-contamination. Spectra were recorded using manufacturer software (UV-Vis analyst), with baseline subtraction applied automatically, and absorbance values exported at 1 nm intervals for analysis. Quality control included periodic blank re-measurements to verify lamp stability, cuvette integrity checks, and ambient temperature maintenance (25 ± 1˚C) throughout the process.

## O-PTIR spectrometry

Water and urine samples were digested with 10% (v/v) hydrogen peroxide (SigmaAldrich, St. Louis, MO, USA) at 60◦C for 24 h. The digests were sieved through a 13 µm stainless-steel mesh, and the filtrates were vacuum-filtered onto a 400 nm goldcoated polycarbonate membrane (Sterlich Corporation, WA, USA). Each membrane was affixed to a glass slide and analyzed on an mIRage IR microscope (Photothermal Spectroscopy Corp., Santa Barbara, CA, USA) equipped with a quantum-cascade midIR pulsed tunable laser and a 532 nm detection laser (Proble Power: 4.4%; IR Power 24%). O-PTIR spectra were collected from 941 to 3007 $cm^{-1}$ in reflectance mode with two averages per point using PTIR Studio 4.4. The Feature Finder module was used to locate and measure the size of discrete particles. Spectra were matched against an internal polymer library (Hit Quality Index; HQI > 0.7) to confirm polymer identity.

## SEM Examination Procedures

The SEM used is a Zeiss Sigma 300 VP, provided by the University Central Facilities (UCF) at The University of Hong Kong. The gold coater used is a Bal-tec SCD 005, also from the UCF at The University of Hong Kong.

Sample preparation procedures: 1. Dilution and drying: Dilute the sample with deionized water and drop it onto the surface of a silicon wafer (Si fragment). Allow it to air-dry. 2. Sample mounting: Attach the dried sample to the SEM sample stub using conductive tape (N-type doped silicon is electrically conductive). 3. Conductive treatment: For non-conductive materials, apply a gold (Au) coating using the Bal-tec SCD 005 sputter coater to enhance sample conductivity.

SEM Scanning Parameters: The accelerating voltage (EHT) is 2 kV. The rationale for using a low voltage is as follows: 1. Even with gold coating, the sample's conductivity is still lower than that of conductive materials, and higher voltages may cause electron accumulation and localized overexposure in the images. 2. High voltage could melt organic samples if the electrons fail to dissipate promptly. The working distances (WD) are optimized according to the sample characteristics. Specific values are provided in the Supplemental Method S5.

## MAE, $R^2$, RMSE, and RCV calculation and evaluation methods

We calculated the MAE [52, 53], RMSE [54], and $R^2$ [55] to evaluate the profiling fidelities from multiple perspectives, thereby providing a well-rounded assessment of the model's performance.

MAE reflects the average deviation between predicted and actual values [56]. It is defined as

$$MAE = \frac{1}{n}\sum_{i=1}^{n} |y_i - \hat{y}_i|, \tag{16}$$

where $n$ is the number of samples, $y_i$ is the actual value, and $\hat{y}_i$ is the model's predicted value.

RMSE accounts for the square of errors, making it more sensitive to large deviations [57]. It is computed as

$$RMSE = \sqrt{\frac{1}{n}\sum_{i=1}^{n} (y_i - \hat{y}_i)^2}. \tag{17}$$

$R^2$ measures how well the model explains the overall variance in the data [55]. Its value typically ranges from 0 to 1, but it may be less than 0 if the model performs poorly. The formula is

$$R^2 = 1 - \frac{\sum_{i=1}^{n} (y_i - \hat{y}_i)^2}{\sum_{i=1}^{n} (y_i - \bar{y})^2}, \tag{18}$$

where $\bar{y}$ is the mean of the actual values.

Since MAE uses absolute values, it gives less weight to outliers and emphasizes overall average deviation; its unit is the same as the original target's unit. RMSE squares the residuals, making it more sensitive to outliers or large errors; its unit also matches that of the target variable. In the $R^2$ calculation, the numerator is the sum of squared residuals, and the denominator is the total variance of the data with respect to the mean. A value of $R^2$ close to 1 indicates strong explanatory power and high predictive accuracy. For a more comprehensive evaluation of model performance and suitability, we employ all three metrics simultaneously.

RCV is an enhancement of the traditional coefficient of variation (CV) that leverages outlier-insensitive (i.e., more robust) statistical measures to quantify data dispersion [58]. RCV can be calculated by

$$RCV = \frac{k \times M(|x_i - M(x)|)}{M(x)} \tag{19}$$

where $x = \{x_1, x_2, x_3, ..., x_n\}$ is the sample data, $M(\cdot)$ is the median of the dataset, $k$ is the robust scale, we use $k \approx 1.4826$ for asymptotic normal consistency.

### Noise quantification

The noise quantification for the sample images is performed by first extracting the high-frequency noise components through high-pass filtering, followed by using a statistical measure (standard deviation) to quantify the noise level. Specifically, a Gaussian blur is applied to the grayscale image using a specified kernel size (ksize) and standard deviation (σ). The blurred image is then subtracted from the original grayscale image to obtain a difference image, which primarily contains the original image's high-frequency content (i.e., noise and fine details).

Next, the standard deviation of the difference image is calculated and used as the noise level. A larger standard deviation indicates a higher noise intensity. Since noise generally manifests as high-frequency random fluctuations in the image, this method effectively estimates the noise intensity. Detailed code for estimating noise levels can be found in the Supplemental Method S6.

## Supplementary information

Supplemental Methods S1-S8

Figs. S1 to S47

Tables S1 to S4

Data S1 to S43

Video S1

## Acknowledgements

We would like to thank Prof. George Barbastathis (Department of Mechanical Engineering, Massachusetts Institute of Technology) for helpful discussion of the method, Dr. Hau Kwan Abby Lo and Dr. Chok Hang Yeung (Department of Civil Engineering, The University of Hong Kong) for nature sample collection and sampling information providing, Mr. Rongzhou Chen (Department of Electrical and Electronic Engineering, The University of Hong Kong) for experimental sample preparation, and all the Imaging Systems Lab (The University of Hong Kong) for manuscript review.

## Declarations

• Funding: The work is supported in part by the Research Grants Council of Hong Kong (GRF 17201620, RIF R7003-21) and HKU Seed Fund for Basic Research for New Staff

(2501251073).

• Competing interests: There are no competing interests to declare.

• Ethics approval and consent to participate: Not applicable

• Consent for publication: Not applicable

• Data availability: Authors can confirm that all relevant data are included in the paper and/ or its supplementary information files.

• Materials availability: Not applicable

• Code availability: The code for DSH implementation and conducting the NP signature-based characterization is publicly available on Github (https://github.com/ymzhu19eee/deep-speckle-holography).

• Author contribution: Y. Z. design the method, design the experiment, conduct the experiment, collect the data, write the manuscript, review the manuscript, Y. L. design the experiment, review the manuscript, J. C. collect the data, conduct the experiment, write the manuscript, review the manuscript, D. Y. H. collect the data, write the manuscript, C. W. design the method, write the manuscript, review the manuscript, Y. Z. design the method, analyze the data, write the manuscript, review the manuscript, J. K. F. provide the experimental instrumental source, X. W. conduct the experiment, write the manuscript, F. C. C. provide the experimental resource, B. L. collect the data, write the manuscript, L. F. T. design the method, provide the supervision, E. Y. L. design the method, provide the supervision, get the fund source, review the manuscript.

# References

[1] V. D. Hodoroaba, U. Wolfgang, S. Alexander, eds. *Characterization of nanoparticles: Measurement processes for nanoparticles*. Elsevier, (2019).

[2] M. M. Modena, B. Rühle, T. P. Burg, S. Wuttke, Nanoparticle characterization: What to measure?. *Adv. Mater.*, **31**, 1901556, (2019).

[3] Y. Cheng, Single-particle cryo-EM—How did it get here and where will it go. *Science*, **361**, 876-880, (2018)

[4] H. J. Chung, C. M. Castro, H. Im, H. Lee, R. Weissleder, A magneto-DNA nanoparticle system for rapid detection and phenotyping of bacteria. *Nat. Nanotechnol.*, **8**, 369-375 (2013).

[5] S. A. MacParland, K. M. Tsoi, B. Ouyang, X. Z. Ma, J. Manuel, A. Fawaz, M. A. Ostrowski, B. A. Alman, A. Zilman, W. C. W. Chan, I. D. McGilvray, Phenotype determines nanoparticle uptake by human macrophages from liver and blood. *ACS Nano*, **11**, 2428-2443 (2017).

[6] M. Poudineh, P. M. Aldridge, S. Ahmed, B. J. Green, L. Kermanshah, V. Nguyen, C. Tu, R. M. Mohamadi, R. K. Nam, A. Hansen, S. S. Sridhar, A. Finelli, N. E. Fleshner, A. M. Joshua, E. H. Sargent, S. O. Kelley, Tracking the dynamics of circulating tumour cell phenotypes using nanoparticle-mediated magnetic ranking. *Nat. Nanotechnol.*, **12**, 274-281, (2017).

[7] R. M. Davis, B. Kiss, D. R. Trivedi, T. J. Metzner, J.C. Liao, S. S. Gambhir, Surface-enhanced Raman scattering nanoparticles for multiplexed imaging of bladder cancer tissue permeability and molecular phenotype. *ACS Nano*, **12**, 9669-9679, (2018).

[8] L.F. Tadesse, *Diagnostic Raman Spectroscopy: from Liquid Phase Bacterial Identification to Co-Incubation Free Antibiotic Susceptibility Testing*. Stanford University, (2021).

[9] Y. S. S. Yang, P. U. Atukorale, K. D. Moynihan, A. Bekdemir, K. Rakhra, L. Tang, F. Stellacci, D. J. Irvine, High-throughput quantitation of inorganic nanoparticle biodistribution at the single-cell level using mass cytometry. *Nat. Commun.*, **8**, 14069 (2017).

[10] S. R. LaPlante, V. Roux, F. Shahout, G. LaPlante, S. Woo, M. M. Denk, S. T. Larda, Y. Ayotte, Probing the free-state solution behavior of drugs and their tendencies to self-aggregate into nano-entities. *Nat. Protoc.*, **16**, 5250-5273 (2021).

[10] Di Silvio, D, *Physical-chemical characterization of nanoparticles in relevant biological environments and their interactions with the cell surface* (Doctoral dissertation, University of East Anglia) (2015).

[11] D. Docter, U. Distler, W. Storck, J. Kuharev, D. Wünsch, A. Hahlbrock, S. K. Knauer, S. Tenzer, R. H. Stauber, Quantitative profiling of the protein coronas that form around nanoparticles. *Nature Protoc.*, **9**, 2030-2044 (2014).

[12] S. Winzen, S. Schoettler, G. Baier, C. Rosenauer, V. Mailaender, K. Landfester, K. Mohr, Complementary analysis of the hard and soft protein corona: Sample preparation critically effects corona composition. *Nanoscale*, **7**, 2992-3001 (2015).

[13] B. Kastner, N. Fischer, M. M. Golas, B. Sander, P. Dube, D. Boehringer, K. Hartmuth, J. Deckert, F. Hauer, E. Wolf, H. Uchtenhagen, H. Urlaub, F. Herzog, J. M. Peters, D. Poerschke, R. Lührmann, H. Stark, (2008). GraFix: Sample preparation for single-particle electron cryomicroscopy. *Nat. Methods*, **5**, 53-55 (2008).

[14] P. R. Dunkley, P. E. Jarvie, P. J. Robinson, A rapid Percoll gradient procedure for preparation of synaptosomes. *Nat. Protoc.*, **3**, 1718-1728 (2008).

[15] J. Rodriguez-Loya, M. Lerma, J. L. Gardea-Torresdey, Dynamic light scattering and its application to control nanoparticle aggregation in colloidal systems: A review. *Micromachines*, **15**, 24 (2023).

[16] J. S. Du, *Complex nanoparticle systems: Structures, structure–property relationships, and dynamics* (Doctoral dissertation, Northwestern University) (2021).

[17] W. Schärtl, *Light scattering from polymer solutions and nanoparticle dispersions*. Berlin, Heidelberg: Springer Berlin Heidelberg, (2007).

[18] W. Haiss, N. T. Thanh, J. Aveyard, D. G. Fernig, Determination of size and concentration of gold nanoparticles from UV− Vis spectra. *Anal. Chem.*, **79**, 4215-4221 (2007).

[19] N. T. Le, T. J. Boskovic, M. M. Allard, K. E. Nick, S. R. Kwon, C. C. Perry, Gold nanostar characterization by nanoparticle tracking analysis. *ACS Omega*, **7**, 44677-44688 (2022).

[20] A. Lopez-Serrano Oliver, A. Haase, A. Peddinghaus, D. Wittke, N. Jakubowski, A. Luch, A. Grützkau, S. Baumgart, Mass cytometry enabling absolute and fast quantification of silver nanoparticle uptake at the single cell level. *Anal. Chem.*, **91**, 11514-11519 (2019).

[21] Dykman, M. I. Large fluctuations and fluctuational transitions in systems driven by colored Gaussian noise: A high-frequency noise. *Physical Review A*, 42(4), 2020 (1990)

[22] Mazzoleni, P., Matta, F., Zappa, E., Sutton, M. A., & Cigada, A. Gaussian pre-filtering for uncertainty minimization in digital image correlation using numerically-designed speckle patterns. *Optics and Lasers in Engineering*, 66, 19-33 (2015)

[23] Commoner, B., Lipkin, D.: The application of the Beer-Lambert law to optically anisotropic systems. *Science*, 110(2845), 41-43 (1949)

[24] Swinehart, D. F.: The beer-lambert law. *Journal of Chemical Education*, 39(7), 333, (1962)

[25] E. J. Cho, H. Holback, K. C. Liu, S. A. Abouelmagd, J. Park, Y. Yeo, Nanoparticle characterization: State of the art, challenges, and emerging technologies. *Mol. Pharm.*, **10**, 2093-2110 (2013).

[26] P. M. Carvalho, M. R. Felício, N. C. Santos, S. Gonçalves, M. M. Domingues, Application of light scattering techniques to nanoparticle characterization and development. *Front. Chem.*, **6**, 237 (2018).

[27] H. Cao, A. P. Mosk, S. Roter, (2022). Shaping the propagation of light in complex media. *Nat. Phys.*, **18**, 994-1007 (2022).

[28] J. W. Goodman, *Introduction to Fourier Optics*, Fourth edition edn. W.H.Freeman, Macmillan Learning, New York (2017).

[29] W. Haiss, N. T. Thanh, J. Aveyard, D. G. Fernig, Determination of size and concentration of gold nanoparticles from UV− Vis spectra. *Anal. Chem.*, **79**, 4215-4221 (2007).

[30] J. Liu, K. E. Murphy, R. I. MacCuspie, M. R. Winchester, Capabilities of single particle inductively coupled plasma mass spectrometry for the size measurement of nanoparticles: A case study on gold nanoparticles. *Anal. Chem.*, **86**, 3405-3414 (2014).

[31] P. Zhang, M. Cao, A. J. Chetwynd, K. Faserl, F. Abdolahpur Monikh, W. Zhang, R. Ramautar, L. A. Ellis, H. H. Davoudi, K. Reilly, R. Cai, K. E. Wheeler, D. S. T. Martinez, Z. Guo, C. Chen, I. Lynch, I. Analysis of nanomaterial biocoronas in biological and environmental surroundings. *Nat. Protoc.*, **19**, 3000-3047 (2024).

[32] A. A. Ashkarran, Decentralized nanoparticle protein corona analysis may misconduct biomarker discovery. *Nano Today*, **62**, 102745 (2025).

[33] S. Lee, J. Cha, S. Shin, M. Yoon, G. Sánchez-Velázquez, S. Yang, M. S. Strano, S. Y. Cho, Corona Dynamics of Nanoparticles and Their Functional Design Space in Molecular Sensing. *ACS Nano*, *19*(27), 24653-24668 (2025).

[34] D. Luo, C. Li, X. Bai, Y. Shi, R. Wang, Photoaging-induced variations in heteroaggregation of nanoplastics and suspended sediments in aquatic environments: A case study on nanopolystyrene. *Water Res.*, **268**, 122762 (2025).

[35] Y. Zhu, Y. Li, J. Huang, Y. Zhang, Y. W. Ho, J. K. H Fang, E. Y. Lam, Advanced optical imaging technologies for microplastics identification: Progress and challenges. *Adv. Photon. Res.* **5**, 2400038 (2024).

[36] J. Su, J. Ruan, D. Luo, J. Wang, Z. Huang, X. Yang, Y. Zhang, Q. Zeng, Y. Li, W. Huang, L. Cui, C. Chen, Differential photoaging effects on colored nanoplastics in aquatic environments: physicochemical properties and aggregation kinetics. *Environ. Sci. & Technol.*, **57**, 15656-15666 (2023).

[37] D. Kazakov, T. P. Letsou, M. Piccardo, L. L. Columbo, M. Brambilla, F. Prati, S. Dal Cin, M. Beiser, N. Opačak, P. Ratra, M. Pushkarsky, D. Caffey, T. Day, L. A. Lugiato, B. Schwarz, F. Capasso, Driven bright solitons on a mid-infrared laser chip. *Nature*, **641**, 1–7 (2025).

[38] B. G. Oripov, D. S. Rampini, J. Allmaras, M. D. Shaw, S. W. Nam, B. Korzh, A. N. McCaughan, A superconducting nanowire single-photon camera with 400,000 pixels. *Nature* **622**, 730–734 (2023).

[39] D. Gehrig, D. Scaramuzza, Low-latency automotive vision with event cameras. *Nature* **629**, 1034–1040 (2024).

[40] C. Wang, S. Zhu, P. Zhang, K. Wang, J. Huang, E. Y. Lam, Angle-based neuromorphic wave normal sensing. *Laser Photonics Rev.* **19**, 2400647 (2025).

[41] P. Zhang, S. Zhu, C. Wang, Y. Zhao, E. Y. Lam, Neuromorphic imaging with super-resolution. *IEEE Trans. Circuits Syst. Video Technol*. **35**, 1715-1727 (2024).

[42] S. Chen, Y. Li, Y. Wang, H. Chen, A. Ozcan, Optical generative models. *Nature* **644**, 903–911 (2025).

[43] B. E. Saleh, M.C. Teich, *Fundamentals of Photonics,* **2**, John Wiley& Sons, (2019).

[44] R. Brown, Absorption and scattering of light by small particles. Optica Acta:International *J. Opt.* **31**, 3–3 (1984).

[45] J. W. Goodman, *Statistical Optics*.John Wiley & Sons, (2015).

[46] Y. Sun, X. Zuo, S. K. Sankaranarayanan, S. Peng, B. Narayanan, G. Kamath, Quantitative 3D evolution of colloidal nanoparticle oxidation in solution. *Science* **356**, 303–307 (2017).

[47] A. S. Sznitman, *Brownian Motion, Obstacles and Random Media.* Springer, Verlag Berlin Heidelber (2013).

[48] G. E. Karniadakis, I. G. Kevrekidis, L. Lu, P. Perdikaris, S. Wang, L. Yang, Physics-informed machine learning. *Nat. Rev. Phys.* **3**, 422–440 (2021).

[49] M. Horie, N. Mitsume, Physics-embedded neural networks: Graph neural pde solvers with mixed boundary conditions. *Adv. Neural. Inf. Process. Syst.* **35**, 23218–23229 (2022).

[50] R. Guo, Z. Lin, T. Shan, X. Song, M. Li, F. Yang, S. Xu, A. Abubakar, A.: Physics

embedded deep neural network for solving full-wave inverse scattering problems. *IEEE Trans. Antennas Propag.* **70**, 6148–6159 (2021).

[51] D. F. Swinehart, The beer-lambert law. *J. of Chem. Educ.* **39**, 333 (1962).

[52] K. Elsayed, Mean absolute deviation: Analysis and applications. *Int. J. Bus. Stat. Anal.* **2** (2015).

[53] J. Bell, *Machine Learning: Hands-on for Developers and Technical Professionals*. John Wiley & Sons, Indiana (2020).

[54] J. Fan, R. Li, C. H. Zhang, H. Zou, *Statistical Foundations of Data Science*. Chapman and Hall/CRC, London (2020).

[55] S. Nakagawa, H. Schielzeth, A general and simple method for obtaining $R^2$ from generalized linear mixed-effects models. *Methods Ecol. Evol.* **4**, 133–142 (2013).

[56] C. J. Willmott, K. Matsuura, Advantages of the mean absolute error (MAE) over the root mean square error (RMSE) in assessing average model performance. *Clim. Res.* **30**, 79–82 (2005).

[57] T. O. Hodson, Root mean square error (RMSE) or mean absolute error (MAE): When to use them or not. *Geosci. Model Dev.,* **15**, 1–10 (2022).

[58] Z. Botta-Dukát, Quartile coefficient of variation is more robust than CV for traits calculated as a ratio. *Sci. Rep.* **13**, 4671 (2023).